\documentclass[letterpaper,twocolumn,10pt]{article}
\usepackage{usenix2019_v3}

\usepackage{tikz}
\usepackage{amsmath}
\usepackage{hyperref}

\begin{document}

\date{}

\title{\Large \bf A Simple and Agile Cloud Infrastructure to Support Cybersecurity Oriented Machine Learning Workflows }

\author{
{\rm Ajay Lakshminarayanarao}\\
Sophos AI \\
ajay.lakshminarayanarao@sophos.com
\and
{\rm Konstantin Berlin}\\
Sophos AI \\
konstantin.berlin@sophos.com
} 

\maketitle

\begin{abstract}
Generating up to date, well labeled datasets for machine learning (ML) security models is a unique engineering challenge, as large data volumes, complexity of labeling, and constant concept drift makes it difficult to generate effective training datasets. Here we describe a simple, resilient cloud infrastructure for generating ML training and testing datasets, that has enhanced the speed at which our team is able to research and keep in production a multitude of security ML models.
\end{abstract}

\section{Introduction}
One of the best-known secrets of machine learning (ML) is that the most reliable way to get more accurate models is by getting more training data and more accurate labels \cite{halevy2009unreasonable}. Unfortunately, generating larger, more relevant datasets is arguably a bigger challenge in the security domain than in most other domains, due to two major complications. The first is that labeling information is usually not available at time of observation, but slowly evolves over time (days to months) as more information is observed. The second complication is the constant concept drift in the in-the-field distributions, as well as accuracy of the labeling algorithm itself. In such an environment, developing a model around a specific ``gold'' dataset is not enough, and any deployed model is constantly retrained on newer data, with constantly evolving labeling strategies, that needs to be propagated to all observables immediately.

\begin{figure*}
\begin{center}
\includegraphics[width=0.9\textwidth, angle=0]{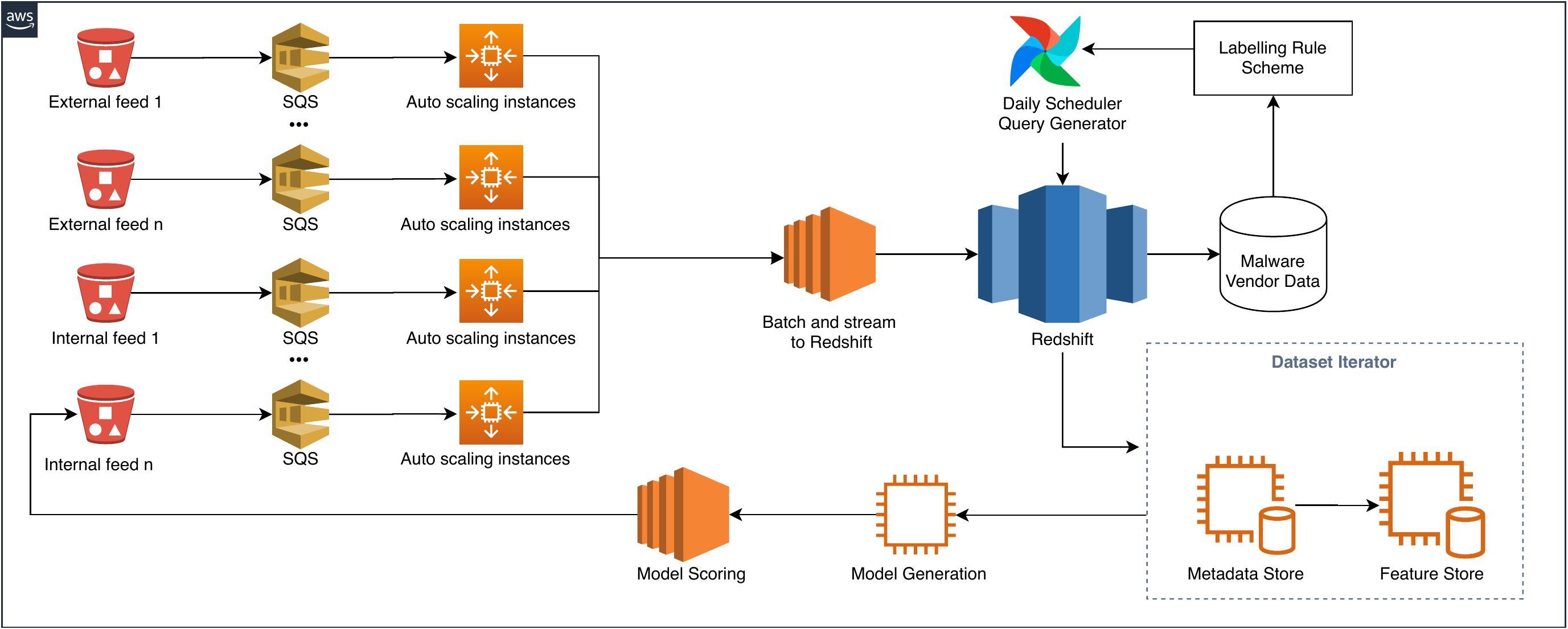}
\end{center}
\caption{\label{fig:architecture} Diagram of the described cloud infrastructure and training loop. The top left of the diagram shows how all incoming data is batched, preprocessed, and inserted into Redshift. The top right of the diagram shows how internal Redshift queries are used to generate metadata around stored data. The bottom of the diagram illustrates how data is exported into a ``Dataset'' abstraction, which is used to drive model training. Incoming data is scored by ML models, and the results are inserted back into Redshift. }
\end{figure*}

Here we describe an AWS cloud infrastructure that we have developed to enable rapid prototyping and retraining of machine learning security models (see Figure \ref{fig:architecture}), consisting of an ingestion workflow, a data warehousing solution, and a training/test dataset abstraction interface. The central tenants of this infrastructure design are, low-cost, easy maintenance, and agility around dataset generation and labeling. This architecture encompasses auto scaling ingestion of internal and external feeds, data warehousing, computing on-demand malware labels, extracting metadata and feature generation for research and periodic deployment of machine learning models.

\section{Ingestion Workflow}
 The ingestion workflow is a generic configurable daemon responsible for fetching internal and external feeds for standard ETL processes. These daemons perform the minimum amount of parsing and flattening of incoming data, as well as batching, such that the data can be efficiently inserted into a database or stored in S3 \cite{s3}. The output batch is formed when the desired time limit is reached, or the batch size reaches a certain size. The end state for any specific message is typically either a data warehouse, in the case of metadata, or S3 storage, in case of feature vectors and binary blobs.

The main advantage of implementing a custom batch based solution is flexibility in how the data is processed, and control of underlying compute resources, which can drive down costs. The advantage of using SQS \cite{sqs}, a managed messaging service, is low-cost, high scalability, and fault tolerance. An individual ingestion feed's response time of ingestion can vary between a minute to hours, depending on desired batch size.

\section{Warehousing}
Redshift \cite{gupta2015amazon}, a columnar storage based data warehouse is the critical analytic engine that helps us execute multiple essential tasks. It acts as a storage and query engine for large amounts of indexed data, as a compute engine for aggregating malware event scans, and as a provider of training data by generating metadata and labels periodically.

A large fraction of data transformation and cleaning is performed in Redshift. This results in very wide input tables, however, because of columnar storage, we are able to store all information at low-cost, are able to recompute full data transformations and cleaning sequences and select only the required columns using SQL.

Once the data is cleaned, we execute nightly aggregations that join the data across tables, presenting a final, cleaned, consistent view of all the information to the end users. This workflow is highly agile, since we can re-slice and dice almost the entirety of all known metadata by simply modifying a SQL query. Using SQL, and not complex ingestion pipelines to transform data made it possible for our ML researchers to have significant control over the training data and labels, without having to engage with outside teams, or having to write complicated processing pipelines.

\section{Labeling}

By far, the most difficult aspect of deploying security ML models is the constant need to have new and up to date labels on as much of the data as possible. Sources for labeling can be outside intelligence feeds, like VirusTotal \cite{vt}, internal automated systems (ex. sandboxes and scanrigs), as well as manual human labeling. One of the main goals for our infrastructure, is to ingest this firehose of feeds, and combine the various feeds together to maximize labeling coverage, as well as increase reliability of any given label.

The major advantage of our Redshift centric approach is that we are able to control the labeling logic directly using SQL statements. Thus, it is pretty trivial to adjust the labeling logic, and to immediately propagate to all observables. No additional compute resources are required, and it results in drastically simplifying and democratizing the labeling process. 

\section{Dataset}
Machine learning models, specifically deep learning models, require frequent slicing and randomly ordered iteration of training and test datasets. All our model training is done through an abstract software layer called ``Dataset'', which provides a uniform interface for slicing and iterating training and test datasets. Simple wrappers are created to interface between Dataset and common ML libraries like Tensorflow \cite{abadi2016tensorflow} and Scikit-learn \cite{pedregosa2011scikit}.

Dataset is storage independent, and is defined by a set of iterators around a metadata store and a feature store, and a set of collectors that populate the Dataset from Redshift, S3 and other storage services.

\section{Conclusion}

We presented a simple end-to-end cloud infrastructure that enables our team of ML researchers to quickly research and productize new and improved ML models. This infrastructure scales to very large ingestion volumes, while requiring only a small maintenance team and a limited budget. While cost was our primary consideration when we deployed it, it turned out that the most impactful long-term aspect of our system was actually enabling our researchers to directly control the critical aspects of ML workflows, like labeling, dataset cleaning and selection, through simple SQL queries.

\bibliographystyle{plain}
\bibliography{references}

\end{document}